\documentclass[a4paper,11pt]{article}

\usepackage{jheppub} 

\usepackage[T1]{fontenc} 

\usepackage{bm}
\usepackage{slashed}
\usepackage{graphicx}
\usepackage{amsmath,amssymb,amsfonts,graphicx}
\usepackage{dsfont, mathrsfs}
\usepackage{bm}
\usepackage{slashed}
\usepackage[svgnames]{xcolor}
\usepackage{placeins}
\usepackage{natbib,hyperref}  
\usepackage{url}

\newcommand{\Eq}[1]{Eq.~(\ref{#1})}
\newcommand{\al}[1]{

\begin{align} #1 \end{align}}
\newcommand{\non}{\nonumber}
\newcommand{\vect}[1]{\boldsymbol{#1}}
\newcommand{\vh}[1]{\boldsymbol{\hat{#1}}}

\def\Tr{\mathrm{Tr}}

\def\SA{\sphericalangle}

\def\fS{\varphi_S}
\def\fR{\varphi_R}

\def\fL{\varphi_L}

\def\fK{\varphi_{k}}

\def \fk{\varphi_{k}}
\def \fkB{\varphi_{\bar{k}}}

\def\fKR{\varphi_{KR}}

\def \fq{\varphi_{q}}

\def\kT{\vect{k}_T}
\def\kBT{\vect{\bar{k}}_T}
\def\RT{\vect{R}_T}

\def\RBT{\vect{\bar{R}}_T}
\def\qT{\vect{q}_T}

\def\PP{\vect{P}_\perp}

\def\PLP{\vect{P}_{\Lambda\perp}}

\def\ST{\vect{S}_T}

\newcommand{\SLa}{\vect{S}_{\Lambda}}
\newcommand{\sLa}{\vect{s}_{\Lambda}}
\newcommand{\lLa}{s_{L}}

\newcommand{\llangle}{\left\langle}
\newcommand{\rrangle}{\right\rangle}
\title{\boldmath Semi-inclusive back-to-back  production of a hadron pair and a single hadron in $e^+e^-$ annihilation} 

\preprint{ADP-18-21/T1069}

\author[a,1]{Hrayr~H.~Matevosyan,\note{ORCID: http://orcid.org/0000-0002-4074-7411}}
\author[b,c,2]{Aram~Kotzinian,\note{ORCID: http://orcid.org/0000-0001-8326-3284}}
\author[a,3]{Anthony~W.~Thomas\note{ORCID: http://orcid.org/0000-0003-0026-499X}}

\affiliation[a]{ARC Centre of Excellence for Particle Physics at the Terascale,\\ 
and CSSM, Department of Physics, \\
The University of Adelaide, Adelaide SA 5005, Australia
\\ http://www.physics.adelaide.edu.au/cssm
}

\affiliation[b]{Yerevan Physics Institute,
2 Alikhanyan Brothers St.,
375036 Yerevan, Armenia
}

\affiliation[c]{INFN, Sezione di Torino, 10125 Torino, Italy
}

\emailAdd{hrayr.matevosyan@adelaide.edu.au}
\emailAdd{aram.kotzinian@cern.ch}
\emailAdd{anthony.thomas@adelaide.edu.au}

\abstract{
Inclusive hadron production in $e^+e^-$ annihilation has long been used to study both single hadron fragmentation functions (FF) and dihadron fragmentation functions (DiFF).  In particular, the polarized DiFFs can be accessed in electron-positron annihilation by measuring azimuthal correlations between two back-to-back pairs of hadrons in the center of mass system, where the relevant structure functions can be  expressed as convolutions of two (polarized) DiFFs. Here we explore the advantages of measuring the inclusive back-to-back production of a single hadron on one side against a hadron pair on the opposite side of the detector in two jet events. The leading twist cross section for this process contains convolutions of the corresponding single hadron FFs on one side and the DiFFs for the hadron pair on the other side, which furnishes several interesting new opportunities. A measurement of the unpolarized cross section with a number of different types of observed hadrons will help in untangling the quark flavor dependence of the unpolarized DiFFs, when the results are analyzed together with the inclusive measurements of dihadron pairs, such as those recently performed by the {\tt BELLE}  collaboration. Even more interesting, with a polarized hyperon on one side we can study the quark spin-dependent DiFFs of an unpolarized hadron pair on the other side. This, in turn, will allow us to test the universality of the spin-dependent DiFFs entering the cross sections of  electron-positron annihilation and semi-inclusive deep inelastic scattering processes.
}


\date{\today}                     

\begin{document} 
\maketitle
\flushbottom

\section{Introduction}
\label{SEC_INTRO}

The study of the hadronization process, which is quantified by various fragmentation functions, has gained a great deal of attention in recent years~\cite{Metz:2016swz}. This has been motivated by the ability of the new experiments to measure various azimuthal correlations in the deep inelastic scattering processes that involve polarized FFs and DiFFs. For example, the measurements of the quark transverse polarization dependent Collins FF and the so-called interference DiFF (IFF) in electron-positron annihilation in {\tt BELLE}~\cite{Abe:2005zx,Seidl:2008xc,Vossen:2011fk,Seidl:2017qhp} and  {\tt BaBar}~\cite{TheBABAR:2013yha} experiments, have allowed the extraction of the quark transversity parton distribution function (PDF)~\cite{Kang:2015msa,Anselmino:2015sxa,Bacchetta:2011ip,Pisano:2015wnq} using semi-inclusive deep inelastic scattering (SIDIS) measurements with one and two detected final state hadrons by the {\tt HERMES}~\cite{Airapetian:2004tw,Airapetian:2008sk} and {\tt COMPASS}~\cite{Adolph:2012nw,Adolph:2014fjw} collaborations.\hspace{-0.1cm} It is worth mentioning that the cross section for hadron production in $e^+e^-$ annihilation involves a summation over all intermediate quark-antiquark pairs, thus requir- ing additional input in order to access the quark flavor dependence of the FFs and DiFFs.

The key for such a combined analysis is the universality of the FFs and DiFFs entering the cross section of both electron-positron annihilation and SIDIS processes, which was proven explicitly for the transverse-momentum dependent FFs~\cite{Meissner:2008yf,Gamberg:2010uw}, while similar arguments should apply in the case of DiFFs~\cite{Bacchetta:2003vn}. This is in contrast to the prediction of the process dependence of the naive-time-reversal-odd  (T-odd) Sivers PDF, which is predicted to change sign between SIDIS and Drell-Yan annihilation~\cite{Collins:2002kn}. An experimental test of such universality for the naive-time-reversal-odd FFs has been proposed in Ref.~\cite{Boer:2010ya}, by exploring the so-called ``polarizing" FF $D_{1T}^\perp$ of a spin $1/2$ hyperon. The ``polarizing" FF describes the correlation of the transverse polarization of a produced hadron with its own transverse momentum in the quark fragmentation process and is a chiral-even function. Thus, in SIDIS with a final state hyperon detected, it couples to the well-determined unpolarized PDF, while in $e^+e^-$ annihilation with back-to-back hyperon and unpolarized hadron production, it couples to the corresponding unpolarized FF. 

Recently, we proposed two new measurements~\cite{Matevosyan:2017liq} of the quark helicity dependent DiFF $G_1^\perp$, both in SIDIS with two final state detected hadrons and two back-to-back hadron pair production in $e^+e^-$ annihilation. The chiral-even nature of $G_1^\perp$ entails that in SIDIS it couples to the well-determined quark unpolarized and helicity PDFs~\cite{Bacchetta:2002ux}. In $e^+e^-$ annihilation though, the  $G_1^\perp$ from one pair couples to that for the second pair on the other side~\cite{Boer:2003ya,Matevosyan:2018icf}, making it not possible (or at least very hard) to determine the sign of this function. Thus, such measurements allow one to test only for the agreement between the magnitudes of $G_1^\perp$ extracted from the two measurements.

 In this work we propose a new measurement in $e^+e^-$ annihilation, exploiting the production of a single inclusive hadron back-to-back to a hadron pair, where the relevant cross section should involve convolutions of FFs for the single hadron in one jet and the DiFFs for the hadron pair produced in the opposite jet. The purpose of such a measurement is two-fold, and will leverage our knowledge of the single hadron FFs. First, the absolute cross section measurements will provide a wider basis for extracting the quark flavor dependence of the DiFFs, especially when analyzed together with the inclusive hadron pair measurements in the same jet~\cite{Seidl:2017qhp}. Secondly, by studying various azimuthal asymmetries we can better determine the polarized DiFFs, and also access their sign.
 
 This paper is organized in the following way.  In the next section we detail the derivation of the cross section for the proposed new process. In Sec.~\ref{SEC_DIFF_ASYMM}, we explore the relevant unpolarized measurements and the azimuthal asymmetries for accessing the DiFFs. In Sec.~\ref{SEC_LAMBDA}, we examine the particular case where one detects a polarized $\Lambda^0$ hyperon. Finally, we present the summary of our findings and conclusions in Sec.~\ref{SEC_CONCLUSIONS}.

\vspace{-0.2cm}
\section{The cross section calculation}
\label{SEC_XSEC}

In this section we detail the derivation of the leading twist cross section for the process $e^+e^-\to h_1 h_2 +\Lambda + X$, where the electron and positron with momenta $l$ and $l'$ annihilate into an intermediate virtual photon with momentum $q=l+l'$. In the final state, we detect a hadron $\Lambda$ with momentum $P_\Lambda$, produced back-to-back to the unpolarized hadron pair $h_1,h_2$ of momenta $P_1$ and $P_2$, as illustrated in Fig.~\ref{PLOT_EE_KINEMATICS}. Here $\Lambda$ denotes either an unpolarized hadron ( for example $\pi, K$, etc ) hadron, or a spin $1/2$ baryon withe a polarization vector $\SLa$. We restrict our consideration to the case where the center-of-mass energy of the electron-positron pair is far below the mass of the $Z$ boson.  We use the conventional framework for the inclusive hadron production in $e^+e^-$ annihilation~\cite{Boer:1997mf,Boer:1997qn,Boer:2003ya, Matevosyan:2018icf}.  In the next subsection we first describe the kinematics of the process and then detail the calculation  of the cross section in the following subsection.

\subsection{Kinematics}
\label{SUBSEC_EE_KINEMATICS}

\begin{figure}[htb]
\centering 
\includegraphics[width=1\columnwidth]{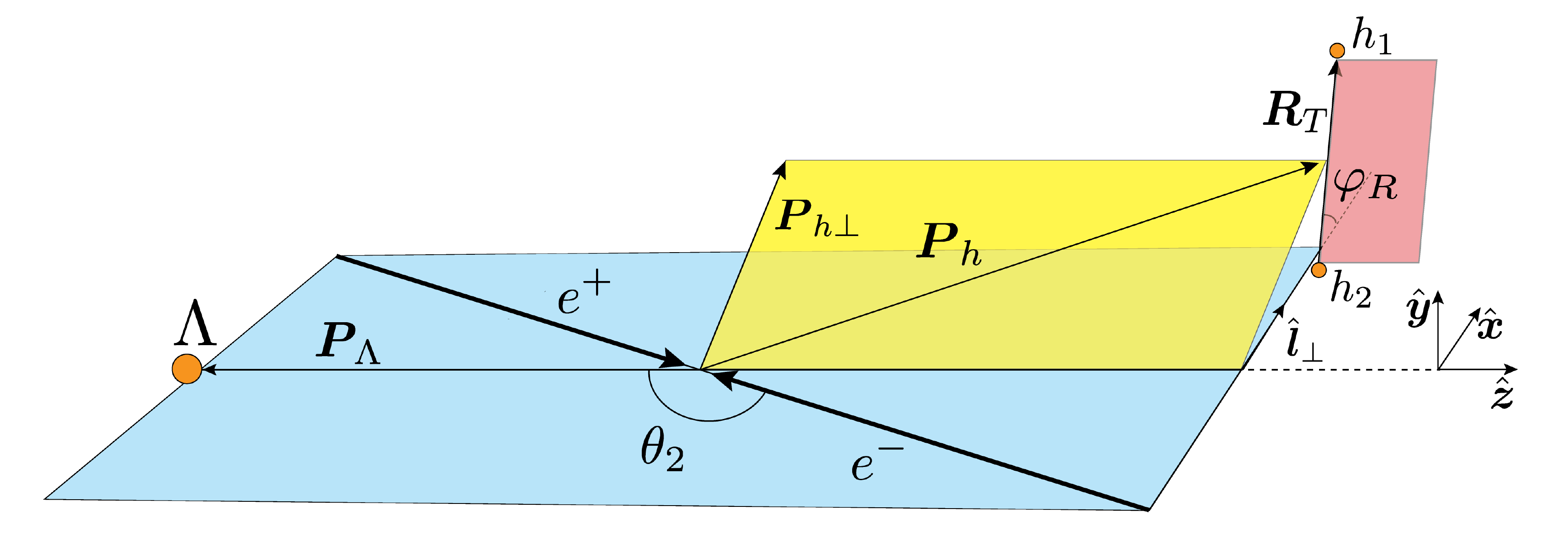}
\caption{The kinematics of $e^+e^-\to h_1 h_2 +\Lambda + X$ process.}
\label{PLOT_EE_KINEMATICS}
\end{figure}

  For our calculations we choose the $e^+e^-$ center-of-mass coordinate system, where the $\hat{z}$ axis is chosen to point opposite to the $\Lambda$'s 3-momentum $\vect{P}_\Lambda$. We denote the components of 3-vectors transverse to $\hat{z}$ axis with subscript $_\perp$. The $\hat{x}$ axis is then taken along the transverse  component $\vect{l}_\perp$ of the electron momentum.  Together, the $\hat{z}$ and $\hat{x}$ axes span the lepton plane. In dihadron studies it is useful to replace $P_1, P_2$ by the total and the relative hadron momenta, defined as
\al
{
\label{EQ_P_TOT}
 P_h &= P_1 + P_2,
\\
\label{EQ_P_REL}
 R &= \frac{1}{2}( P_1 - P_2),
}
and the invariant mass of the pair is defined by $P_h^2 = M_h^2$.
  
  We consider the kinematic regime where the virtuality of the time like momentum of the virtual photon is much larger than the typical hadronic mass scales, thus defining $Q^2\equiv q^2$ we can ignore any contributions of order $1/Q$. At the QED and QCD leading order approximation, the virtual photon produces a quark and an antiquark pair $e^+e^-\to \gamma^* \to q + \bar{q}$, which then hadronize and produce two back-to-back jets. Further, we assume that we are in the "leading hadron approximation", where a significant fraction of the energy in each jet is carried by the observed hadrons, that is $P_h \cdot P_\Lambda \sim Q^2$. We can use these two large momenta to define light-like directions to conveniently decompose the hadronic tensor, as done earlier for other reactions~\cite{Boer:1997mf, Boer:2003ya}.  For this, we choose a coordinate system where the transverse components of the light-cone momenta\footnote{The light-cone components of a 4-vector $a$ are defined as $a = (a^+,a^-,\vect{a}_T)$, where $a^\pm = \frac{1}{\sqrt{2}}(a^0 \pm a^3)$}  of $P_h$, $P_\Lambda$ vanish, $\vect{P}_{hT}=0$ and $\vect{P}_{\Lambda T}=0$. Introducing two light-like vectors $n_+$ and $n_-$, such that $n_+^2 = n_-^2=0$ and $n_+ n_-=1$, we can decompose the relevant large momenta as
\al
{
\label{EQ_PH_LC}
 P_h^\mu &= \frac{M_h^2}{z_h Q \sqrt{2}} n_-^\mu + \frac{z_h Q}{\sqrt{2}} n_{+}^\mu  \approx \frac{z_h Q}{\sqrt{2}} n_+^\mu,
 \\
 \label{EQ_PHB_LC}
P_\Lambda^\mu &= \frac{z_\Lambda Q}{\sqrt{2}} n_{-}^\mu + \frac{M_\Lambda^2}{z_\Lambda Q \sqrt{2}} n_+^\mu \approx \frac{z_\Lambda Q}{\sqrt{2}} n_-^\mu,
 \\
 \label{EQ_Q_LC}
 q^\mu &= \frac{ Q}{\sqrt{2}} n_{-}^\mu + \frac{Q}{\sqrt{2}} n_+^\mu  + q_T^\mu,
 }
where $M_\Lambda$ is the mass of $\Lambda$, and the momentum fractions are defined as
\vspace{-0.5cm}
\al
{
 z_h &= \frac{2 P_h\cdot q}{Q^2} \approx  \frac{P_h^+}{q^+} \equiv z,
 \\
 z_\Lambda &= \frac{2 P_\Lambda \cdot q}{Q^2} \approx \frac{P_\Lambda^-}{q^-} \equiv  \bar{z},
}
and 
\vspace{-0.5cm}
\al
{
  - q_T^2 = Q_T^2 \ll Q^2.
}

  In this system, for any 4-vector,  we can obtain the components orthogonal to  $n_\pm$ using
\vspace{-0.5cm}
\al
{
g_T^{\mu \nu} &= g^{\mu\nu} - n_+^\mu n_-^\nu - n_+^\nu n_-^\mu,
\\
\epsilon_T^{\mu \nu} &=  \epsilon^{\mu \nu \rho \sigma} n_{+ \rho} n_{- \sigma},
}
where $g^{\mu\nu}$ is the metric tensor and we use the convention $\epsilon^{0123} = +1$.

 The leptonic tensor, on the other hand, is usually evaluated in the coordinate system depicted in Fig.~\ref{PLOT_EE_KINEMATICS}, where $\vect{P}_\Lambda = \vect{q}_\perp =0$. In this system, we can again define normalized time-like and space like vectors along the dominant momenta
 \vspace{-0.5cm}
 \al
{
 \hat{t} &= \frac{q}{Q},
 \\
 \hat{v} &= 2 \frac{P_\Lambda}{\bar{z} Q} - \hat{t} \, .
}
The projections of the 4-vectors onto the two orthogonal $_\perp$ directions can be obtained using the tensors
\vspace{-0.5cm}
\al
{
g_\perp^{\mu\nu} &= g^{\mu\nu} - \hat{t}^\mu \hat{t}^\nu + \hat{v}^\nu \hat{v}^\mu,
\\
\epsilon_\perp^{\mu \nu} &= - \epsilon^{\mu \nu \rho \sigma}  \hat{t}_\rho \hat{v}_\sigma.
}

 Interestingly, the two perpendicular projection tensors  differ only by a factor of order $Q_T/Q$, since
\vspace{-0.5cm}
\al
{
g_\perp^{\mu\nu} &= g_T^{\mu\nu} - \frac{n_+^\mu q_T^\nu + n_+^\nu q_T^\mu}{Q} \, .
}
Thus we can neglect  the differences between the $_T$ and $_\perp$ components of the vectors in this work.

\subsection{Cross section}
\label{SUBSEC_EE_XSEC}

The cross section for this process is given by the convolution of the leptonic and hadronic tensors
\al
{
 \frac{2 P_1^0 2P_2^0 2 P_\Lambda^0 d \sigma}
 {d^3 \vect{P}_1  d^3 \vect{P}_2  d^3 \vect{P}_\Lambda  }
 = \frac{\alpha_{em}^2}{Q^6} L_{\mu\nu} W^{\mu\nu}_{(3h)},
}
where $L_{\mu\nu}$ is the leptonic tensor, $W^{\mu\nu}_{(3h)}$ is the hadronic tensor, and $\alpha_{em}$ is the fine structure constant. The leptonic tensor can be evaluated using standard methods (e.g.~\cite{Boer:1997mf}), yielding
\vspace{-1cm}
\al
{
\label{EQ_LEPT_TENS}
 L_{\mu\nu} = Q^2\Bigg[ 
-2 A(y) g_\perp^{\mu\nu} 
+ 4B(y) \hat{v}^\mu \hat{v}^\nu &- 4B(y)\Big( \hat{l}_\perp^\mu \hat{l}_\perp^\nu + \frac{1}{2} g_\perp^{\mu\nu} \Big)
\\ \non 
&
- 2C(y) B^{1/2}(y) \Big( \hat{v}^\mu \hat{l}_\perp^\nu + \hat{v}^\nu \hat{l}_\perp^\mu \Big) 
 \Bigg],
}
where the normalized perpendicular part of the leptons's 4-momentum is defined as
\vspace{-0.5cm}
\al
{
 \hat{l}_\perp^\mu = \frac{l_\perp^\mu}{|\vect{l}_\perp|}.
}
The usual coefficient functions are given by
\vspace{-0.5cm}
\al
{
 A(y) &= \frac{1}{2} - y + y^2,
\\
B(y) &= y(1-y),
\\
C(y) &= 1-2y,
}
where the lepton momentum fractions are defined as
\vspace{-0.5cm}
\al
{
\label{EQ_Y_TH2}
 y = \frac{P_h \cdot l}{P_h \cdot q} \approx \frac{l^-}{q^-} = \frac{1+ \cos \theta_2}{2}.
}
Here  $\theta_2$ is the angle between the 3-momentum of the electron $l$ and the $\vh{z}$ axis in the center-of-mass frame.

The phase space factor can be then written in the conventional notation
\vspace{-0.5cm}
\al
{
d^9V
&\equiv
\frac {d^3 \vect{P}_1  d^3 \vect{P}_2  d^3 \vect{P}_\Lambda  } {2 P_1^0 2P_2^0 2 P_\Lambda^0 }
=
\frac{d^2 \vect{P}_{h\perp} d z \ d\fR d M_h^2  d \xi}{8 z}
\frac{  Q^2  \bar{z} d \bar{z} d\Omega_\Lambda } {8 }
,
}
where we used $\vect{P}_{\Lambda\perp}=0$ and the leading order relations in the hard scale $Q$
\vspace{-0.5cm}
\al
{&
  P_h^- \ll P_h^+ , \rightarrow E_h \sim \frac{1}{\sqrt{2}} P_h^+,
  \\
  &
  R^- \ll R^+ , \rightarrow E_R \sim \frac{1}{\sqrt{2}} R^+,
  \\
  &
  \frac{R^+}{P_h^+} = \xi - 1/2 \sim \frac{E_R}{E_h},
}
and we replaced $|\vect{R}_T|$ with the invariant mass of the hadron pair $M_h$ using the relation
\vspace{-0.5cm}
\al
{
 \vect{R}_T^2 & =  \xi (1-\xi) M_h^2 - M_1^2 (1-\xi) - M_2^2 \xi.
}

 The spherical angle of the $\Lambda$ hadron can be expressed in terms of $y$ and the azimuthal angle of $\vect{l}_\perp$ as
\al
{
  d\Omega_\Lambda = 2 d y  d\fL,
}
yielding
\vspace{-0.9cm}
\al
{
 d^9V &= \frac{Q^2}{32} \frac{\bar{z}}{z}
d^2 \vect{P}_{h\perp} d z \ d\fR \ d M_h^2 \ d \xi
\ d \bar{z} \  d y \ d \fL
\\\
&= z \bar{z}  \frac{Q^2}{32}
d^2 \qT \ d z \ d\fR \ d M_h^2 \ d \xi
\ d \bar{z} \  d y \ d \fL \, ,
}
where in the last expression we substituted $\vect{P}_{h\perp}$ with the transverse component of the intermediate virtual photon's 3-momentum in the light-cone frame $\qT = - \vect{P}_{h\perp}/z$.

The hadronic tensor is defined as
\vspace{-0.5cm} 
\al
{
W^{\mu\nu}_{(3h)}(q; P_h, R, P_\Lambda) 
 = \frac{1}{(2\pi)^{7}}\sum_X  \int  &\frac{d^3 \vect{P}_X}{(2\pi)^3 2 P_X^0} 
 \ (2\pi)^4 \  \delta(q - P_X - P_h - P_\Lambda )
\\ \non
& \times \langle 0| J^\mu(0) | P_X; P_h, R, P_\Lambda \rangle
 \langle  P_X; P_h, R, P_\Lambda | J^\nu(0) | 0 \rangle \, .
}

Using the parton picture, we can decompose the hadronic tensor in terms of the quark-quark correlators $\Delta$ and $\bar{\Delta}$ for the production of a hadron pair and a hyperon in the fragmentation of the quark and the antiquark with momenta $k$ and $\bar{k}$, respectively
\vspace{-0.5cm}
\al
{
\label{EQ_W_TENS_D_DBAR}
 W^{\mu\nu}_{(3h)} \approx & \ \frac{3}{(2\pi)^3}
 \sum_{a} e_a^2
 \int d^4 k \ d^4 \bar{k}\  \delta^4(q - k - \bar{k})
 \Tr \Bigg[ \bar{\Delta}(\bar{k}, P_\Lambda) \gamma^\mu 
\Delta(k, P_h, R) \gamma^\mu
\Bigg],
}
where $a$ and $e_a$ denote the flavor and the fractional electric charge of the fragmenting quark, while the pre factor is the number of quark colors $N_c=3$. We can then  integrate the $k^+$ and $\bar{k}^-$ components using the $\delta$ function and rewrite the hadronic tensor as a convolution of the conventional integrated quark-quark correlators
\vspace{-0.5cm}
\al
{
\label{EQ_W_TENS_INT_D_DBAR}
 W^{\mu\nu}_{(3h)} \approx  \ \frac{3 (32 z) (4\bar{z})}{(2\pi)^3}
 \sum_{a} e_a^2
 \int d^2 \kT &d^2 \kBT \ \delta^2(\qT - \kT - \kBT)
\\ \non
&\times
 \Tr \Bigg[ \bar{\Delta}(\bar{z}, \kBT)_{\bar{k}^- = P_\Lambda^-/\bar{z}} \gamma^\mu 
\Delta(z,\xi, \kT, \RT)_{k^+ = P_h^+/z} \gamma^\mu
\Bigg],
}
where the light-cone momentum fractions for the dihadron correlators are defined in terms of those for the individual hadrons,
\vspace{-1cm}
\al
{
 z &= z_1  + z_2,
 \\
 \xi &= \frac{z_1}{z}  = 1- \frac{z_2}{z} \, ,
}
where $z_i = {P_i^+}/{k^+}$ are the light-cone momentum fractions of each hadron in the pair.

The two-hadron fragmentation of a quark is described by a quark-quark correlator~\cite{Bianconi:1999cd, Radici:2001na, Boer:2003ya, Boer:2008fr}
\vspace{-0.8cm}
\al
{
\Delta_{ij}(k; P_h, R)
 = \sum_X \int d^4 \zeta &e^{i k\cdot \zeta} \langle 0 |\psi_i(\zeta) | P_h R, X \rangle \langle P_h R, X | \bar{\psi}_j(0) | 0 \rangle,
 }
which, for the case of an unpolarized hadron pair and at the leading twist approximation, is parametrized  via four DiFFs
\vspace{-0.5cm}
\al
{
\frac{1}{32z} \int d k^- \Delta(k, &P_h, R) |_{k^+ = P_h^+/z} \equiv \Delta(z, \xi,\kT, \RT) 
\\ \non
 =\frac{1}{4\pi}\, \frac{1}{4}\Bigg\{
 & D_1 \slashed{n}_{+}
  - G_1^\perp\frac{\epsilon_{\mu\nu\rho\sigma} \gamma^\mu n_+^\nu k_T^\rho R_T^\sigma}{M_h^2} \gamma_5
 + H_1^\SA\frac{\sigma_{\mu\nu} R_T^\mu n_+^\nu}{M_h}
 + H_1^\perp\frac{\sigma_{\mu\nu} k_T^\mu n_+^\nu}{M_h}
 \Bigg\}.
}
Here $D_1$ is the unpolarized DiFF,  while the helicity-dependent DiFF, $G_1^\perp$, describes the correlation of the fragmenting quark's longitudinal polarization with the vector product of the transverse momenta of the two hadrons. The two remaining DiFFs, $H_1^\SA$  (IFF) and $H_1^\perp$, describe the correlations between the quark transverse polarization with the relative and total transverse momenta of the pair, respectively. The DiFFs here are defined in a frame where their total momentum 3-momentum is along the $\hat{z}$ axis, and we can choose as two transverse vectors $\RT$ and $\kT$, which are the transverse components of $R$ and $k$ in this system. We can relate $\kT = - \PP/ z $ to the transverse component of $P$ in a frame where $k$ has no transverse component. In general,  DiFFs depend on $z, \xi , \kT, \RT$. It is worth noting, that only $\kT\cdot \RT$ correlations are possible for our scalar functions and so the azimuthal dependence of the DiFFs can be  expanded in a Fourier cosine series
\vspace{-0.cm}
\al
{
 \label{EQ_FOURIER_FRK}
D_1 (z, \xi, \kT^2, \RT^2,& \cos(\fKR)) 
  = \frac{1}{\pi} \sum_{n=0}^\infty
\frac{\cos(n \cdot \fKR)}{1+\delta^{0,n}} \ D_1^{[n]}(z, \xi, |\kT|, |\RT| ) \, ,
}
where $\fKR \equiv \fK - \fR$ is the relative azimuthal angle between $\kT$ and $\RT$.

For single polarized hadron production the quark-quark correlator now depends on the momentum of the fragmenting quark $k$, and the momentum $P_\Lambda$ and polarization $\SLa$  of the produced hadron. Once again we can integrate over the large component of the light-cone momentum
\vspace{-0.cm}
\al
{
\frac{1}{4z} \int d k^+ \Delta(k, P_\Lambda, \SLa)_{k^- = P_\Lambda^- /z}  \equiv \Delta(z, \kT, \SLa) 
\, ,
}
where $\kT$ is the transverse component of the fragmenting quark's momentum in a system with the $\hat{z}$ axis pointed along the 3-momentum $\vect{P}_\Lambda$. Following the notation in Ref.~\cite{Mulders:1995dh}, we can expand the correlator in terms of the relevant leading-twist transverse momentum dependent (TMD) FFs
\vspace{-0.cm}
\al
{
\Delta(z, \kT, \SLa)  = \frac{1}{4} \Bigg\{&
\quad D_1 \slashed{n}_{+}
 + D_{1T}^\perp \frac{\epsilon_{\mu\nu\rho\sigma} \gamma^\mu n_+^\nu k_T^\rho S_T^\sigma}{M_h}
\\ \non &
 - \Big(\lambda_\Lambda G_{1L} + G_{1T} \frac{\kT \cdot \ST}{M_h} \Big) \slashed{n}_{+} \gamma_5
 - H_{1T}\  {i \sigma_{\mu\nu} S_T^\mu n_{+}^\nu} \gamma_5
\\ \non & 
 -   \Big(\lambda_\Lambda  H_{1L}^\perp + H_{1T}^\perp \frac{\kT \cdot \ST}{M_h} \Big) {i \sigma_{\mu\nu} k_T^\mu n_{+}^\nu} \gamma_5
 + H_{1}^\perp \frac{ \sigma_{\mu\nu} k_T^\mu n_{+}^\nu}{M_h}
\Bigg\},
}
where $\lambda_\Lambda$ and $\ST$ are the helicity and the transverse polarization vectors of the produced $\Lambda$ hadron. The FFs are functions of $z$ and $\PLP^2$.  Here $\PLP = - z \kT$ is the transverse momentum of the produced hadron in a system where the 3-momentum of the quark $\vect{k}$ has no transverse component.

\newpage
Finally, we can calculate the cross section
\newcommand{\vS}{2.3cm}
\al
{
\label{EQ_XSEC_V1}
&\frac{d \sigma \Big( e^+e^- \to (h_1 h_2) + \Lambda + X \Big)}{d^2 \qT \ d z \ d\fR \ d M_h^2 \ d \xi \ d \bar{z} \  d y } 
 =\frac{3 \alpha_{em}^2}{(2\pi)^2 Q^2} z^2 \bar{z}^2 \sum_{a} e_a^2 
\\ \non
& \hspace{1.7cm}
\times
\Bigg\{
\qquad
A(y)\ \mathcal{F}\Bigg[  D_1^{a\to h_1 h_2} D_1^{\bar{a}\to \Lambda}\Bigg]
\\ \non
& \hspace{\vS} - S_T A(y)\ \mathcal{F}\Bigg[\frac{\bar{k}_T}{M_\Lambda} \sin(\fkB - \fS)\  D_1^{a\to h_1 h_2} D_{1T}^{\perp,\bar{a}\to \Lambda} \Bigg]
\\ \non
& \hspace{\vS} + \lambda_\Lambda A(y)\ \mathcal{F}\Bigg[ \frac{k_T R_T}{M_h^2} \sin(\fk - \fR)\  G_1^{\perp,a\to h_1 h_2} G_{1L}^{\bar{a}\to \Lambda} \Bigg]
\\ \non
& \hspace{\vS} + S_T A(y)\ \mathcal{F}\Bigg[\frac{k_T R_T}{M_h^2} \sin(\fk - \fR) \frac{\bar{k}_T}{M_\Lambda} \cos(\fkB - \fS)\  G_1^{\perp,a\to h_1 h_2} G_{1T}^{\bar{a}\to \Lambda} \Bigg]
\\ \non
& \hspace{\vS} + S_T B(y)\ \mathcal{F}
\Bigg[\Big(\ \ \ \frac{k_T}{M_h}\sin(\fK + \fS) H_1^{\perp, a\to h_1 h_2}
\\ \non
& \hspace{4.8cm} 
+ \frac{R_T}{M_h} \sin(\fR + \fS) H_1^{\SA, a\to h_1 h_2} \Big) \  H_{1T}^{\bar{a}\to \Lambda} \Bigg]
\\ \non
& \hspace{\vS} + \lambda_\Lambda B(y)\ \mathcal{F}
\Bigg[\Big(\ \ \ \frac{k_T}{M_h} \sin(\fK + \fkB) H_1^{\perp, a\to h_1 h_2}
\\ \non
& \hspace{4.8cm} 
+ \frac{R_T}{M_h}  \sin(\fR + \fkB) H_1^{\SA, a\to h_1 h_2}\Big)\  \frac{\bar{k}_T}{M_\Lambda}    H_{1L}^{\perp, \bar{a}\to \Lambda} \Bigg]
\\ \non
& \hspace{\vS} + S_T B(y)\ 
\mathcal{F}\Bigg[\Big(\ \ \ \frac{k_T}{M_h} \sin(\fK + \fkB) H_1^{\perp, a\to h_1 h_2}
\\ \non
& \hspace{4.9cm}  
+ \frac{R_T}{M_h}\sin(\fR + \fkB)  H_1^{\SA, a\to h_1 h_2}  \Big) \ \frac{\bar{k}_T^2}{M_\Lambda^2}  \cos(\fkB-\fS) H_{1T}^{\perp, \bar{a}\to \Lambda} 
\Bigg]
\\ \non
& \hspace{\vS} +\hspace{0.45cm} B(y)\ \mathcal{F}
\Bigg[\Big(\ \ \ \frac{k_T}{M_h} \cos(\fK + \fkB)H_1^{\perp, a\to h_1 h_2}
\\ \non
&\hspace{4.8cm}
+ \frac{R_T}{M_h} \cos(\fR + \fkB)   H_1^{\SA, a\to h_1 h_2} \Big)\ \frac{\bar{k}_T}{M_\Lambda}  H_{1}^{\perp, \bar{a}\to \Lambda} \Bigg]
\quad \Bigg\}  \, ,
}
where $\fS$ is the azimuthal angle of $\ST$ and we have integrated over the trivial dependence on the azimuthal angle of the lepton plane $\fL$. The transverse momentum convolution, $\mathcal{F}$, is defined as
\al
{
\label{EQ_F_CONVOL}
\mathcal{F}&[w D^a \bar{D}^{\bar{a}} ] 
\\ \non
&=  \int d^2 \kT d^2 \kBT
   \delta^2(\vect{k}_T + \vect{\bar{k}}_T - \vect{q}_T)w( \kT, \kBT, \RT, \RBT)
 D^a(z, \xi, \kT^2, \RT^2, \kT \cdot \RT  )
 D^{\bar{a}}(\bar{z},  \kBT^2) \, .
}

Integrating the cross section over over $\qT$, $\fR$ and $\xi$ yields the unpolarized "collinear" cross section
\vspace{-0.5cm}
\al
{
\label{EQ_XSEC_INT_UPOL}
 \frac{d \sigma \Big( e^+e^- \to(h_1 h_2) + \Lambda + X \Big)}{d z  \ d M_h^2 \ d \bar{z} \  d y }
& =  \frac{3 \alpha_{em}^2}{(2\pi)^2 Q^2}  A(y) \sum_{a} e_a^2 
\ D_1^{a\to h_1 h_2}(z, M_h^2) \ \bar{D}_1^{\bar{a} \to \Lambda}(\bar{z}),
}
where
\vspace{-0.5cm}
\al
{
 D_1^a(z, M_h^2) &= z^2 \int d^2 \kT \int d\xi \int d\fR \, D_1^a (z, \xi, \kT^2, \RT^2, \kT \cdot \RT) 
 \\ \non
  &=  z^2 \int d^2 \kT \int d\xi \, D_1^{a,[0]} (z, \xi, |\kT|, |\RT|)  ,
}
and
\vspace{-0.5cm}
\al
{
D_1^{ \bar{a} }(z) = \bar{z}^2 \int d^2 \kBT D_1^{\bar{a}}(\bar{z}, \kBT^2).
}

Further integrating the result in \Eq{EQ_XSEC_INT_UPOL} over $M_h^2$ we recover the same expression as for two back-to-back hadron production in Eq.(77) of Ref.~\cite{Boer:1997mf}. This is natural, as the integrated DIFF
\vspace{-0.5cm}
\al
{
D_1^{a\to h_1 h_2}(z) \equiv \int d M_h^2 \ D_1^{a\to h_1 h_2}(z, M_h^2),
}
can be thought of describing the collinear fragmentation into a "parent" particle that carries the total light-cone momentum fraction $z$ of the pair $h_1 h_2$.

The cross sections of the measurements of a hadron pair in the same hemisphere, such as those recently performed by the {\tt BELLE} collaboration~\cite{Seidl:2015lla,Seidl:2017qhp}, involve a sum over all the fragmenting quark flavors. Thus, leveraging the relatively well-determined single hadron FFs in a combined analysis of the single hadron pair cross section with that in \Eq{EQ_XSEC_INT_UPOL} would enable the flavor separation of the unpolarized DiFFs. 

For example, for DiFF into $\pi^+\pi^-$ pairs, the isospin and charge symmetries entail
\vspace{-0.5cm}
\al
{
&
 D_1^{u\to \pi^+\pi^-} =  D_1^{\bar{u}\to \pi^+\pi^-} \approx  D_1^{d\to \pi^+\pi^-} = D_1^{\bar{d}\to \pi^+\pi^-},
 \\
& D_1^{s\to \pi^+\pi^-} = D_1^{\bar{s}\to \pi^+\pi^-}.
}
For the one pair inclusive production, omitting the contributions of all the heavier flavors, we can approximate
\vspace{-0.5cm}
\al
{
d \sigma( e^+e^- \to (h_1 h_2) +   X ) \sim \sum_q e_q^2\ D_1^{q\to \pi^+\pi^-} \approx  \frac{5}{9} D_1^{u\to \pi^+\pi^-}(z) + \frac{1}{9} D_1^{s\to \pi^+\pi^-}(z) \, .
}
Hence the separation of the light and strange quark DiFF contributions is not possible without additional input. For the associated hadron production in the opposite hemisphere,  the cross section can be approximated as
\vspace{-0.5cm}
\al
{
\label{EQ_UNP_PIPI_PI}
d \sigma( e^+e^- \to (h_1 h_2) +  \pi^+ + X )
\sim  \frac{5}{9} D_1^{u\to \pi^+\pi^-}(z) D_1^{u^+\to \pi^+}(\bar{z}) + \frac{1}{9} D_1^{s\to \pi^+\pi^-}(z)D_1^{s^+\to \pi^+}(\bar{z}),
}
where
\vspace{-0.5cm}
\al
{
D_1^{q^+\to h}(\bar{z}) \equiv D_1^{q\to h}(\bar{z}) + D_1^{\bar{q}\to h}(\bar{z}).
}

The unpolarized FFs to charged pions have been phenomenologically extracted from experimental measurement by a number of different groups~\cite{deFlorian:2014xna, Hirai:2016loo, Sato:2016wqj,Ethier:2017zbq,Bertone:2017tyb}. Thus the measurement of the cross section in~\Eq{EQ_UNP_PIPI_PI} would allow us to extract the flavor dependence of the DiFF $D_1^{q\to \pi^+\pi^-}$. Similar arguments can be used to explore other hadron channels and combinations.  For example, choosing a $\pi^+K^+$ pair on one side and a $K^-$ hadron on the other side would significantly enhance the "double favored" channel $( \bar{s}\to \pi^+K^+, s \to K^- )$  for moderate to large values of $z_1$, $z_2$, and $\bar{z}$, as can also be seen in hadronization models~\cite{Ito:2009zc,Matevosyan:2011zza,Matevosyan:2011ey,Matevosyan:2013aka}. The contributions of all the other quark flavors would include at least one "unfavored" fragmentation, which will be suppressed for our chosen region of light-cone momentum fractions. Thus, such a  measurement would allow us to extract a DiFF for a specific quark flavor, $\bar{s}\to \pi^+K^+$. In principle, a global fit, using all possible measurements of the single and dihadron fragmentations would yield the best estimates of these functions, with the most realistic uncertainties.

\section{The azimuthal asymmetries}
\label{SEC_DIFF_ASYMM}

In this section we will discuss several asymmetries that give access to various combinations of polarization-dependent FFs and DiFFs, that can be measured at {\tt BELLE} and the upcoming {\tt BELLE~II} experiments, similar to the measurements of the asymmetries for back-to-back hadrons~\cite{Boer:1997nt,Boer:2008fr} and back-to-back hadron pairs~\cite{Boer:2003ya,Matevosyan:2018icf}.

We define the integral of the cross section in \Eq{EQ_XSEC_V1}, weighted with an arbitrary function $\mathcal{I}$, as
\vspace{-0.5cm}
\al
{
\label{EQ_XSEC_AVERAGE}
\langle \mathcal{I} \rangle \equiv
  \int d\xi & \int d\fR \int d^2 \qT  \ \mathcal{I} 
\  \frac{d \sigma \Big( e^+e^- \to (h_1 h_2) + \Lambda + X \Big)}{d^2 \qT \ d z \ d\fR \ d M_h^2 \ d \xi \ d \bar{z} \  d y } ,
}
where the unpolarized integrated cross section in \Eq{EQ_UNP_PIPI_PI}  is simply
\vspace{-0.5cm}
\al
{
\label{EQ_AV_INT_UPOL}
 \frac{d \sigma \Big( e^+e^- \to (h_1 h_2) + \Lambda + X \Big)}{d z  \ d M_h^2 \ d \bar{z} \  d y } = \langle 1 \rangle .
}
%

\subsection{The asymmetry induced by the Collins effect for unpolarized $\Lambda$}

For an unpolarized $\Lambda$, only the first and the last terms in \Eq{EQ_XSEC_V1} contribute. In addition to the unpolarized term, the correlation of the transverse polarizations of the fragmenting quark and antiquark pair is described by the convolution of the Collins function $H_{1}^{\perp, \bar{a}\to \Lambda}$  for a $\Lambda$ with the two dihadron analogues $H_1^{\perp, a\to h_1 h_2}, H_1^{\SA, a\to h_1 h_2}$ for the hadron pair $h_1h_2$. Thus, we can access this term by considering the following weighted average
\al
{
&\llangle \frac{q_T}{M_\Lambda}\cos(\fq+ \fR) \rrangle = \frac{3 \alpha_{em}^2}{(2\pi)^2 Q^2}\frac{B(y)}{M_\Lambda^2 M_h}
\\ \non
& \hspace{1.5cm}\times \sum_{a} e_a^2  \int d\xi  \int d\fR \int d^2 \qT \int d^2 \kT \int d^2 \kBT
\delta^2(\vect{k}_T + \vect{\bar{k}}_T - \vect{q}_T) q_T \cos(\fq + \fR)   
 \\ \non
& \hspace{1.5cm}\times \Bigg[ \Big( k_T \bar{k}_T \cos(\fK + \fkB) H_1^{\perp, a\to h_1 h_2}  
+ R_T \bar{k}_T \cos(\fR + \fkB)   H_1^{\SA, a\to h_1 h_2} \Big)    H_{1}^{\perp, \bar{a}\to \Lambda} \Bigg],
}
where we can use the $\delta$ function to write
\al
{
 \int d^2 \qT  \ \delta^2(\vect{k}_T + \vect{\bar{k}}_T - \vect{q}_T) \ q_T\cos(\fq + \fR)  = (k_T \cos(\fk+ \fR)  + \bar{k}_T \cos(\fkB+ \fR) ).
 }
We further use the Fourier cosine decomposition for the DiFFs in ~\Eq{EQ_FOURIER_FRK} to conclude
\al
{
\label{EQ_AV_COLL}
\llangle \frac{q_T}{M_\Lambda}\cos(\fq+ \fR) \rrangle
=& \frac{3 \alpha_{em}^2}{(2 \pi)^2 Q^2} B(y) \sum_{a} e_a^2 
\  H_1^{\SA, a\to h_1 h_2}(z, M_h^2) \  H_{1}^{\perp \bar{a}, [1]}(\bar{z}),
}
noting that 
\al
{
  H_1^{\SA, a\to h_1 h_2}&(z, M_h^2) 
  \\ \non
 &  \equiv z^2 \int d^2\kT  \int d \xi  \, \Big[ \frac{|\RT|}{M_h}H_1^{\SA, [0]} (z, \xi, |\kT|, |\RT|) + \frac{|\kT|}{M_h}  H_1^{\perp, [1]} (z, \xi, |\kT|, |\RT|) \Big],
}
is the same integrated IFF that enters in SIDIS~\cite{Bacchetta:2003vn} and in $e^+e^-$ two hadron pair~\cite{Matevosyan:2018icf} asymmetries, while
\al
{
H_{1}^{\perp \bar{a}, [1]}(\bar{z}) \equiv 
\bar{z}^2 \int d^2 \kBT \ \frac{\kBT^2}{2 M_\Lambda^2} \ H_{1}^{\perp, \bar{a}\to \Lambda} (z, \kBT^2),
}
is the first moment of the Collins function.

The corresponding weighted azimuthal asymmetry then takes a simple form
\al
{
\label{EQ_ASYMM_COLL}
 A^{Coll} =  \frac{B(y)}{A(y)} \dfrac{\sum_{a} e_a^2\ H_1^{\SA, a\to h_1 h_2}(z, M_h^2) \  H_{1}^{\perp \bar{a}, [1]}(\bar{z}) } {\sum_{a} e_a^2\ D_1^{a\to h_1 h_2}(z, M_h^2) \ \bar{D}_1^{\bar{a} \to \Lambda}(\bar{z}) } \, .
}
%

\subsection{The asymmetries for a longitudinally polarized $\Lambda$}
\label{SUB_SEC_POL}

To devise the measurements involving the polarization of the produced $\Lambda$, it is important to underline that  the cross section expression in \Eq{EQ_XSEC_V1} is a conditional probability of the process for the given polarization $\SLa$. We have no a priory knowledge of this polarization, and it cannot be measured directly on an event-by-event basis. Thus, we need to extract the polarization $\sLa$ that $\Lambda$ acquires in the back-to-back fragmentation process, which can be measured in experiment by considering the average angular distributions of the final state particles. Employing the spin density matrix formalism, we can infer from general considerations~\cite{Berestetsky:1982aq,Mulders:1995dh,Matevosyan:2016fwi}, that the cross section of this process can be expressed as
\vspace{-0.5cm}
\al
{
\label{EQ_XSEC_S}
 \frac{d\sigma}{dV} = \alpha + \vect{\beta} \cdot \SLa  \sim 1+ \SLa \cdot \sLa .
}
Thus, the acquired polarization vector is simply given by
\vspace{-0.5cm}
\al
{
 \sLa = \frac{\vect{\beta}}{\alpha},
}
where $\alpha$ and $\vect{\beta}$ are only functions of the momentum of $\Lambda$ and the polarization of the fragmenting quark. The latter in our work is expressed using the correlations with the momenta of $h_1,h_2$, so that the acquired polarization $\sLa$ only depends on the momenta of the observed particles. Further, it is clear from \Eq{EQ_XSEC_S}, that the average of the acquired polarization over the kinematic variables is given by
\vspace{-0.5cm}
\al
{
 \langle \sLa \rangle = \frac{\langle \vect{\beta} \rangle}{\langle \alpha \rangle},
}
as the probability of the process itself is given by $\alpha$. The coefficients $\alpha$ and $\vect{\beta}$ can be easily read off directly from \Eq{EQ_XSEC_V1}. For example, $\alpha$ corresponds to the first and the last terms on the right hand side of \Eq{EQ_XSEC_V1}, while the longitudinal part $\beta_L$ of the coefficient $\vect{\beta}$ is given by the terms multiplied by $\lambda_\Lambda$. In Sec.~\ref{SEC_LAMBDA} we present an example of a measurement for the longitudinal polarization of $\Lambda^0$ baryon.

Let us consider the correlations for the longitudinally polarized $\Lambda$ that are manifest in two terms in \Eq{EQ_XSEC_V1}. The term that involves $G_1^{\perp,a\to h_1 h_2}$ and $G_{1L}^{\bar{a}\to \Lambda}$ describes the correlations between the longitudinal polarizations of the fragmenting quark and anti-quark. We can access the helicity of $\Lambda$ by considering the weighted average
\vspace{-0.5cm}
\al
{
\label{EQ_AV_G1}
\llangle \beta_L \rrangle _{G_1^\perp G_{1L}} =  \llangle \frac{q_T}{M_h}\sin(\fq - \fR) \rrangle =  \frac{3 \alpha_{em}^2}{(2\pi)^2 Q^2} A(y) \sum_{a} e_a^2 
\ G_1^{\perp,a\to h_1 h_2}(z, M_h^2) \  G_{1L}^{\bar{a}\to \Lambda}(\bar{z}),
}
where
\vspace{-0.5cm}
\al
{
G_1^{\perp,a\to h_1 h_2}&(z, M_h^2)
\\ \non
 &\equiv z^2 \int d^2 \kT \frac{\kT^2}{2 M_h^2} \int d \xi \frac{R_T}{M_h}
\Big( G_1^{\perp,[0]}(z, \xi, |\kT|, |\RT|) - G_1^{\perp,[2]}(z, \xi, |\kT|, |\RT|) \Big),
}
is the same integrated helicity-dependent DiFF as in Refs.~\cite{Matevosyan:2017liq, Matevosyan:2018icf} and
\vspace{-0.5cm}
\al
{
 G_{1L}^{\bar{a}\to \Lambda}(\bar{z}) \equiv \bar{z}^2 \int d^2 \kBT \ G_{1L}^{\bar{a}\to \Lambda}(\bar{z}, \kBT^2) , 
}
is the integrated helicity FF. Thus, the acquired longitudinal polarization $\lLa$ of $\Lambda$ is 
\vspace{-0.5cm}
\al
{
\label{EQ_POL_G1}
 \llangle \lLa\rrangle^{\sin(\fq - \fR)} (z, M_h^2,\bar{z},y) = \dfrac{\sum_{a} e_a^2 
\ G_1^{\perp,a\to h_1 h_2}(z, M_h^2) \  G_{1L}^{\bar{a}\to \Lambda}(\bar{z})
}
{
\sum_{a} e_a^2  
\ D_1^{a\to h_1 h_2}(z, M_h^2) \ \bar{D}_1^{\bar{a} \to \Lambda}(\bar{z})
 },
 }

A second contribution is acquired by the correlation of the transverse polarizations of the quark and the anti quark, which couples the two analogues of Collin DiFFs on one side with the "Kotzinian-Mulders" type FF. The relevant weighting for this term is
\vspace{-0.5cm}
\al
{
\label{EQ_AV_H1L}
\llangle \beta_L \rrangle _{H_1^{\SA} H_{1L}^{\perp}} =  \llangle \frac{q_T}{M_\Lambda}\sin(\fq + \fR) \rrangle = \frac{3 \alpha_{em}^2}{(2 \pi)^2 Q^2} B(y) \sum_{a} e_a^2 
\ H_1^{\SA, a\to h_1 h_2}(z, M_h^2) \  H_{1L}^{\perp \bar{a}, [1]}(\bar{z}),
}
where
\vspace{-0.5cm}
\al
{
H_{1L}^{\perp \bar{a}, [1]}(\bar{z}) \equiv 
\bar{z}^2 \int d^2 \kBT \ \frac{\kBT^2}{2 M_\Lambda^2} \ H_{1L}^{\perp, \bar{a}\to \Lambda} (z, \kBT^2).
}
The corresponding acquired helicity is
\vspace{-0.5cm}
\al
{
 \llangle \lLa\rrangle^{\sin(\fq + \fR)} 
(z, M_h^2,\bar{z},y) = \frac{B(y)}{A(y)}\dfrac{\sum_{a} e_a^2 
\ H_1^{\SA, a\to h_1 h_2}(z, M_h^2) \  H_{1L}^{\perp \bar{a}, [1]}(\bar{z})
}
{
\sum_{a} e_a^2 
\ D_1^{a\to h_1 h_2}(z, M_h^2) \ \bar{D}_1^{\bar{a} \to \Lambda}(\bar{z})
 }.
 }
%

\subsection{The asymmetries for a transversely polarized $\Lambda$}

Finally, we extract the correlations for the transverse polarization of $\Lambda$.  The are  four relevant terms contributing to the transverse part of $\vect{\beta}_T$ in \Eq{EQ_XSEC_V1}. Here we can proceed by considering the $\hat{x}$ and $\hat{y}$ components of this vector, labelled $\beta_x$ and $\beta_y$, respectively.

The transverse polarization acquired by the "polarizing" FF, $D_{1T}^\perp$ mixes with the contribution involving the transversity- and pretzelocity-like FFs of $\Lambda$. 
\vspace{-0.5cm}
\al
{
\label{EQ_AV_DDT}
\llangle \beta_x \rrangle^{\sin(\fq)} =   \llangle {q_T}\sin(\fq) \rrangle 
 = \frac{3 \alpha_{em}^2}{(2\pi)^2 Q^2} \sum_{a} e_a^2 
\Big\{
 &- A(y) {M_\Lambda}D_1^{ a\to h_1 h_2}(z, M_h^2)  \ D_{1T}^{\perp \bar{a}, [1]}(\bar{z})
\\ \non
&+B(y) M_h  H_1^{\perp, a\to h_1 h_2}(z, M_h^2) \  H_{1}^{\bar{a}\to\Lambda}(\bar{z})
\Big\},
}
\al
{
\llangle \beta_y \rrangle^{\cos(\fq)} =   \llangle {q_T}\cos(\fq) \rrangle  
%
 = \frac{3 \alpha_{em}^2}{(2 \pi)^2 Q^2}  \sum_{a} e_a^2 
\Big\{
&  A(y) {M_\Lambda}D_1^{ a\to h_1 h_2}(z, M_h^2)  \ D_{1T}^{\perp \bar{a}, [1]}(\bar{z})
\\ \non
&+B(y) M_h H_1^{\perp, a\to h_1 h_2}(z, M_h^2) \  H_{1}^{\bar{a}\to\Lambda}(\bar{z})
\Big\},
}
where
\vspace{-0.5cm}
\al
{
 H_1^{\perp, a\to h_1 h_2}&(z, M_h^2) 
 \\ \non
 \equiv z^2 & \int d^2\kT  \int d \xi \Big[ \frac{|\kT| |\RT|}{2M_h^2}H_1^{\SA, [1]} (z, \xi, |\kT|, |\RT|) + \frac{|\kT|^2}{2 M_h^2}  H_1^{\perp, [0]} (z, \xi, |\kT|, |\RT|) \Big],
}
and
\vspace{-0.5cm}
\al
 {
D_{1T}^{\perp \bar{a}, [1]}(\bar{z}) &\equiv \bar{z}^2 \int d^2 \kBT \ \frac{\kBT^2}{2 M_\Lambda^2} D_{1T}^{\perp \bar{a}, [1]}(\bar{z}, \kBT^2),
\\ \label{EQ_H1}
H_{1}^{\bar{a}}(\bar{z}) &\equiv \bar{z}^2 \int d^2 \kBT \ \Big(
H_{1T}^{\bar{a}}(\bar{z}, \kBT^2)
 + \frac{\kBT^2}{2 M_\Lambda^2}  H_{1T}^{\perp \bar{a}}(\bar{z}, \kBT^2)
 \Big).
}
This is analogous to the associated $\Lambda$ production asymmetry proposed in Ref.~\cite{Boer:2010ya} and the additional $q_T$ weighting in our case allows us to disentangle the convolution of $\Lambda$ FFs and DiFFs into "collinear" products of the corresponding moments. Here, the contributions involving the chiral-odd  FFs, ignored in~\cite{Boer:2010ya}, enter with different relative sign to those containing the "polarizing" FF. Thus, we can form linear combinations of the two terms to access the contributions from the individual structure functions
\vspace{-0.5cm}
\al
{
\llangle \beta_y \rrangle^{\cos(\fq)} -  \llangle \beta_x \rrangle^{\sin(\fq)}  = 
M_\Lambda \frac{3 \alpha_{em}^2}{2\pi^2 Q^2} A(y)  \sum_{a} e_a^2 
 D_1^{ a\to h_1 h_2}(z, M_h^2)  \ D_{1T}^{\perp \bar{a}, [1]}(\bar{z}),
}
 \al
{
\llangle \beta_y \rrangle^{\cos(\fq)} +  
\llangle \beta_x \rrangle^{\sin(\fq)}  = 
{M_h} \frac{3 \alpha_{em}^2}{2\pi^2  Q^2} B(y) \sum_{a} e_a^2 
 H_1^{\perp, a\to h_1 h_2}(z, M_h^2) \  H_{1}^{\bar{a}\to\Lambda}(\bar{z}),
}
and the corresponding polarizations
\vspace{-0.5cm}
\al
{
\frac{\llangle s_{y} \rrangle^{\cos(\fq)}(z, M_h^2,\bar{z},y)
 - \llangle s_{x} \rrangle^{\sin(\fq)}(z, M_h^2,\bar{z},y) }
 {M_\Lambda}
 = 2 \dfrac{\sum_{a} e_a^2 
 D_1^{ a\to h_1 h_2}(z, M_h^2)  \ D_{1T}^{\perp \bar{a}, [1]}(\bar{z})
}
{
\sum_{a} e_a^2 
 D_1^{a\to h_1 h_2}(z, M_h^2) \ \bar{D}_1^{\bar{a} \to \Lambda}(\bar{z})
 },
}
\al
{
\frac{ \llangle s_{y} \rrangle^{\cos(\fq)}(z, M_h^2,\bar{z},y)
 + \llangle s_{x} \rrangle^{\sin(\fq)}(z, M_h^2,\bar{z},y) }
 {M_h}&
\\ \non
 = & \frac{2B(y)}{A(y)} \dfrac{\sum_{a} e_a^2 
  H_1^{\perp, a\to h_1 h_2}(z, M_h^2) \  H_{1}^{\bar{a}\to\Lambda}(\bar{z})
}
{
\sum_{a} e_a^2 
 D_1^{a\to h_1 h_2}(z, M_h^2) \ \bar{D}_1^{\bar{a} \to \Lambda}(\bar{z})
 }.
}
It is important to stress, that here we are operating with the measured polarizations along the $\hat{x}$ and $\hat{y}$.

The contribution from the term involving $G_1^{\perp, a\to h_1 h_2}$ and another "worm-gear" type FF, $ G_{1T}^{\bar{a}}$, which describes the longitudinal quark antiquark polarization correlations, also admixes with the contributions from several  other terms when calculating the weighted moments, for example with a weight $\llangle {q_T^2} \sin(\fq-\fR) \cos(\fq) \rrangle$. Here, we omit the resulting long and convoluted expression (even by the standards of this manuscript) for brevity.

The last contribution to the transverse polarization can be obtained using weights that only involve the azimuthal angle $\fR$ and no additional factors of $q_T$
\vspace{-0.5cm}
\al
{
\label{EQ_AV_HaHT}
\llangle \vect{\beta}_x \rrangle _{H_1^{\SA} H_{1}}^{\sin(\fR)}= \llangle \vect{\beta}_y \rrangle _{H_1^{\SA} H_{1}}^{\cos(\fR)}= \frac{3 \alpha_{em}^2}{8\pi^2 Q^2} B(y) \sum_{a} e_a^2 
\ H_1^{\SA, a\to h_1 h_2}(z, M_h^2) \  H_{1}^{\bar{a}\to\Lambda}(\bar{z}).
}
The corresponding acquired transverse polarization is
\vspace{-0.5cm}
\al
{
\label{EQ_POL_HaH1}
 \llangle \vect{ s}_{T} \rrangle^{\sin(\fR)}_{x}(z, M_h^2,\bar{z},y) 
 &=  \llangle \vect{ s}_{T} \rrangle^{\cos(\fR)}_{y}(z, M_h^2,\bar{z},y)
 \\ \non
 & = 
 \frac{1}{2}\frac{B(y)}{A(y)}\dfrac{\sum_{a} e_a^2 
\ H_1^{\SA, a\to h_1 h_2}(z, M_h^2) \  H_{1}^{\bar{a}}(\bar{z})
}
{
\sum_{a} e_a^2 \
 D_1^{a\to h_1 h_2}(z, M_h^2) \ \bar{D}_1^{\bar{a} \to \Lambda}(\bar{z})
 }.
 }

This is similar to the IFF asymmetry in SIDIS, where the IFF is multiplied by the integrated transversity PDF. Here, similar to the integrated transversity PDF, the polarized FF $H_1$ is a combination of the TMD "transversity" FF $H_{1T}$  and TMD "pretzelocity" $H_{1T}^\perp$, see \Eq{EQ_H1}.

\section{Treatment of the Polarized $\Lambda^0$}
\label{SEC_LAMBDA}

 In this section we will elaborate the treatment of the measurements involving polarized $\Lambda^0$ production, where either the longitudinal or the transverse polarization is leveraged to access the relevant structure functions in the cross section. The determination of the final state polarization is possible for this hyperon, as it undergoes "self-analyzing" weak decays into a baryon-meson pair. The two hadronic decay channels into $p + \pi^-$ and $n + \pi^0$ have relative branching ratios of $64\%$ and $36\%$, respectively. An important aspect of such decays is the correlation between the polarization vector of the decaying hyperon and the produced hadron's momentum. In particular, taking the $p + \pi^-$ channel as an example,  the decay rate is described as 
 \vspace{-0.5cm}
\al
{
  \frac{dN}{N d\cos{\theta}} \sim 1 + \alpha_{\Lambda} S_\Lambda \cos(\theta),
}
where $S_\Lambda$ is the modulus of the polarization of the $\Lambda^0$, and $\theta$ is the angle between the decay proton momentum and the $\Lambda^0$ polarization vector in its rest frame. For the longitudinal polarization of the $\Lambda^0$ this angle is calculated along the direction of the $\hat z$ axis defined in Section I, Fig.~\ref{PLOT_EE_KINEMATICS}, as we can simply boost along $\hat{z}$ to the $\Lambda^0$ rest frame. The decay parameter $ \alpha_{\Lambda} $ has been measured in a number of experiments in the 60's and 70's~\cite{Cronin:1963zb}, with the Particle Data Group average value of $ \alpha_{\Lambda} =0.642$~\cite{Beringer:1900zz}, which describes the admixture of $s-$ and $p-$ partial waves of the $p, \pi^-$ system~\cite{Lee:1957qs}.

The idea of using $\Lambda^0$'s as a quark longitudinal spin polarimeter in SIDIS and $e^+e^- \to \Lambda^0 +X$ annihilation reactions was originally proposed in Refs.~\cite{Bigi:1976qt, Bigi:1977qe, Augustin:1978wf} and for a quark transverse spin polarimeter in SIDIS in Ref.~\cite{Baldracchini:1980uq}. Since then a number of experimental and phenomenological studies of hyperon polarization in these reactions have been performed, see for examples Refs.~\cite{Buskulic:1996vb, Astier:2000ax, Alekseev:2009ab, Ellis:1995fc, Kotzinian:1997vd, Ellis:2002zv,Ellis:2007ig}. It is worth mentioning that substantial longitudinal polarization of $\Lambda^0$ ($\lambda_\Lambda = -0.320$ for $z > 0.3$) was observed by {\tt ALEPH} Ref.~\cite{Buskulic:1996vb}, indicating the large analyzing power of this measurement. Similarly, the {\tt BELLE} collaboration has recently measured a transverse polarization of the $\Lambda^0$ of order $0.1$ via the so-called "polarizing" fragmentation function Ref.~\cite{Abdesselam:2016nym,Guan:2018ckx}, observing significant analyzing power for the transverse polarization as well.

The important aspect of our proposed measurements is that the polarization of the $\Lambda^0$ couples with the azimuthal angles of the dihadron pair on the other side. On the other hand, the hyperon itself is not directly detected in the experiment, but rather is seen through its decay products. In such a decay, the polarization of the $\Lambda^0$ induces modulations of the angular distributions of the decay products. For example, for the helicity-dependent modulations involving the helicity $\lambda_{\Lambda}$, the polar angle of the decay product $p$ will be modulated by $1+\lambda_{\Lambda} \cos(\theta_p)$, where $\theta_p$ is the polar angle of the proton in  $\Lambda^0$'s rest frame.

 Thus, we need to consider the full observed final state in our process $e^+e^- \to (h_1 h_2) + \Lambda^0 +X \to (h_1 h_2) + (p + \pi^-) +X$.  For extraction of the asymmetry in \Eq{EQ_POL_G1}, we should consider a weighted integral of the complete final state
  \vspace{-0.5cm}
\al
{
\label{EQ_AV_THETA_G1}
  \llangle \cos(\theta_p) \frac{q_T}{M_h}\sin(\fq - \fR) \rrangle   \sim   \alpha_{\Lambda} \
 G_1^{\perp,a\to h_1 h_2} \  G_{1L}^{\bar{a}\to \Lambda},
}
where the measurement of the weighted asymmetries with the detected final state hadrons now involves also the integration over the phase space of the proton. Similar expressions can also be derived for the components of the transverse polarization of  $\Lambda^0$.

Such measurements should be possible with the much improved apparatus of the {\tt BELLE~II} Collaboration in the near future.

\section{Conclusions}
\label{SEC_CONCLUSIONS}

The study of the quark hadronization in $e^+e^-$ experiments is a crucial part of the international efforts to map the 3-dimensional structure of the nucleon and to understand the spin-orbit correlations in the strong interactions. The measurements of the Collins effect and the  two-hadron interference effect at {\tt BELLE}~\cite{Abe:2005zx,Seidl:2008xc,Vossen:2011fk,Seidl:2017qhp} and  {\tt BaBar}~\cite{TheBABAR:2013yha} have been used to extract the nucleon transversity PDF. 

The theoretical underpinning for such measurements of the Collins effect in two back-to-back hadron production was given by Boer and collaborators in Refs.~\cite{Boer:1997mf,Boer:1997qn,Boer:2008fr}, where the relevant asymmetry involves convolution of two Collins FFs for the produced hadrons on each side. A measurement with a polarized hadron on one side and an unpolarized one on the other side was proposed in Ref.~\cite{Boer:2010ya} to test the universality of the T-odd "polarizing" FF  entering the corresponding cross section. Recently, the work by ~\cite{Pitonyak:2013dsu} developed an approach for calculations of the polarized back-to-back hadron pair production in $e^+e^-$ in an arbitrary frame and to  arbitrary precision. The measurement of the spin-dependent DiFF was first proposed in Ref.~\cite{Boer:2003ya} and recently revised and corrected in Ref.~\cite{Matevosyan:2018icf}. Here, the cross section for producing two back-to-back hadron pairs includes convolutions of two DiFFs, one for each pair involved. Note, that the unpolarized FFs can be measured in single inclusive hadron production, while the unpolarized DiFFs are measured by detecting two inclusive hadrons in the same jet~\cite{Seidl:2017qhp}. 

 In this work, we proposed a new inclusive measurement, where an unpolarized hadron pair is detected back-to-back with a single hadron, that may or may not be polarized. We derived the expression for the corresponding cross section in \Eq{EQ_XSEC_V1}, which involves convolutions of the DiFFs for the hadron pair with the FFs for the hadron on the other side. A number of new exciting measurements were then discussed. For example, the measurement of the unpolarized cross section will enable us to determine the flavor dependence of the DiFF using our knowledge of the ordinary unpolarized FFs, which is not possible with the other $e^+e^-$ measurements. Moreover, we can measure a product of the Collins function of a meson with the IFF of the hadron pair~\Eq{EQ_ASYMM_COLL}, that would provide additional information to constrain both functions. This is crucial in improving our knowledge of the transversity PDF from SIDIS measurements involving one and two hadron inclusive final states.
 
The measurement of the hadron polarization dependent asymmetries gives access to a wide variety of combinations of polarized spin $1/2$ baryon FFs and polarized DiFFs, as discussed in Sec.~\ref{SEC_DIFF_ASYMM}. Here we have eight TMD FFs for the baryon convoluted with four DiFFs for the unpolarized hadron pair, whereas in the case of two back-to-back dihadron pairs there are only the four DiFFs involved from each side~\cite{Matevosyan:2018icf}. Nonetheless, further developing the weighted asymmetry method we used in accessing the helicity-dependent DiFFs in Ref.~\cite{Matevosyan:2017liq}, we were able to access the individual combination of TMD FFs with DiFFs, disentangling their transverse momentum convolutions.

  In Sec.~\ref{SUB_SEC_POL} we discussed a critical point for the measurements involving a polarized final state hadron. This polarization is acquired during the hadronization process, in correlation with the momenta of the observed final state particles. Using the spin density matrix formalism, we expressed the polarization vector of $\Lambda$ by utilizing the \Eq{EQ_XSEC_V1}.  Then the measurements of different components of this polarization when analyzed using a particular weighted azimuthal modulations allows us to access specific combinations of FFs with DiFFs. For example, the weighted average in \Eq{EQ_POL_G1} extracts a product of the helicity-dependent DiFF $G_1^\perp$ with the "collinear" helicity FF $G_{1L}$ when measuring the longitudinal polarization of the $\Lambda$. Such a measurement will complement the previously proposed measurements of $G_1^\perp$ in $e^+e^-$ annihilation and SIDIS~\cite{Matevosyan:2017liq, Matevosyan:2018dea}, aimed at testing the universality of $G_1^\perp$.  The transverse polarization-dependent modulation in \Eq{EQ_POL_HaH1} yields a product of the collinear IFF with the collinear "transversity" FF. The last one is strikingly similar to SIDIS, where a similar modulation yields a product of the IFF with the transversity PDF~\cite{Bacchetta:2003vn}. 
  
  Another interesting example is the measurement of the transverse polarization components of the hadron for the weighted azimuthal modulation in \Eq{EQ_AV_DDT}, involving the components of the transverse momentum $\qT$, which yields an admixture of two terms. One is the product of the unpolarized DiFF with the T-odd "polarizing" FF of $\Lambda$, while the second is the product of the Collins-like DiFF with transversity-like FF of $\Lambda$. This is an analogue of the associated production measurement proposed in Ref.~\cite{Boer:2010ya}, where the contributions of the chiral-odd functions (such as Collins function and the transversity-like FF) were neglected. Note, that only the convolution of the chiral-odd unpolarized DiFF with the chiral-odd "polarizing" FF  contribute to the perpendicular components of polarization to $\vect{P}_{h\perp}$, which was considered in Ref.~\cite{Boer:2010ya}. Our results demonstrate that we can separate the  chiral-odd  and chiral-even contributions by forming linear combinations of the $x-$ and $y-$components of the transverse  polarization.
  
The measurement of the baryon polarization was discussed in Sec.~\ref{SEC_LAMBDA}. Such measurements for the $\Lambda^0$ hyperon through the "self-analyzing" weak decays have been long employed as polarimeters. Here, we discussed the relevant weighted asymmetries for the full detected final state particles to measure the relevant structure functions within the "narrow width approximation" for the treatment if the intermediate $\Lambda^0$ baryon, as described for the longitudinal polarization case in \Eq{EQ_AV_THETA_G1}.

This new process completes the set of all possible inclusive measurements with up to two produced hadrons on each side. It presents an exciting new opportunity for exploring the hadronization process using data already collected by the {\tt BELLE} experiment, as well as that to be taken in the upcoming {\tt BELLE~II} experiment. A combined global analysis of these measurements along with the other inclusive hadron production reactions in electron-positron annihilation will allow us to acquire a good understanding of both single hadron and dihadron fragmentations.


\section*{ACKNOWLEDGEMENTS}

We would like to thank Dani\"el~Boer(Groningen) for a useful discussion. The work of H.H.M. and A.W.T. was supported by the Australian Research Council through the ARC Centre of Excellence for Particle Physics at the Terascale (CE110001104) and by the ARC  Discovery Project No. DP151103101, as well as by the University of Adelaide. A.K. was supported by A.I. Alikhanyan National Science Laboratory (YerPhI) Foundation, Yerevan, Armenia. 

\bibliographystyle{JHEP}
\bibliography{fragment}

\providecommand{\href}[2]{#2}\begingroup\raggedright\begin{thebibliography}{10}

\bibitem{Metz:2016swz}
A.~Metz and A.~Vossen, \emph{{Parton Fragmentation Functions}},
  \href{https://doi.org/10.1016/j.ppnp.2016.08.003}{\emph{Prog. Part. Nucl.
  Phys.} {\bfseries 91} (2016) 136}
  [\href{https://arxiv.org/abs/1607.02521}{{\ttfamily 1607.02521}}].

\bibitem{Abe:2005zx}
{\scshape Belle Collaboration} collaboration, R.~Seidl et~al.,
  \emph{{Measurement of azimuthal asymmetries in inclusive production of hadron
  pairs in e+ e- annihilation at Belle}},
  \href{https://doi.org/10.1103/PhysRevLett.96.232002}{\emph{Phys.Rev.Lett.}
  {\bfseries 96} (2006) 232002}
  [\href{https://arxiv.org/abs/hep-ex/0507063}{{\ttfamily hep-ex/0507063}}].

\bibitem{Seidl:2008xc}
{\scshape Belle Collaboration} collaboration, R.~Seidl et~al.,
  \emph{{Measurement of Azimuthal Asymmetries in Inclusive Production of Hadron
  Pairs in e+e- Annihilation at s**(1/2) = 10.58-GeV}},
  \href{https://doi.org/10.1103/PhysRevD.78.032011}{\emph{Phys.Rev.} {\bfseries
  D78} (2008) 032011} [\href{https://arxiv.org/abs/0805.2975}{{\ttfamily
  0805.2975}}].

\bibitem{Vossen:2011fk}
{\scshape Belle Collaboration} collaboration, A.~Vossen et~al.,
  \emph{{Observation of transverse polarization asymmetries of charged pion
  pairs in $e^+e^-$ annihilation near $\sqrt{s}=10.58$ GeV}},
  \href{https://doi.org/10.1103/PhysRevLett.107.072004}{\emph{Phys.Rev.Lett.}
  {\bfseries 107} (2011) 072004}
  [\href{https://arxiv.org/abs/1104.2425}{{\ttfamily 1104.2425}}].

\bibitem{Seidl:2017qhp}
{\scshape Belle} collaboration, R.~Seidl et~al., \emph{{Invariant-mass and
  fractional-energy dependence of inclusive production of di-hadrons in
  $e^+e^-$ annihilation at $\sqrt{s}=$ 10.58 GeV}},
  \href{https://doi.org/10.1103/PhysRevD.96.032005}{\emph{Phys. Rev.}
  {\bfseries D96} (2017) 032005}
  [\href{https://arxiv.org/abs/1706.08348}{{\ttfamily 1706.08348}}].

\bibitem{TheBABAR:2013yha}
{\scshape BaBar} collaboration, J.~P. Lees et~al., \emph{{Measurement of
  Collins asymmetries in inclusive production of charged pion pairs in $e^+e^-$
  annihilation at BABAR}},
  \href{https://doi.org/10.1103/PhysRevD.90.052003}{\emph{Phys. Rev.}
  {\bfseries D90} (2014) 052003}
  [\href{https://arxiv.org/abs/1309.5278}{{\ttfamily 1309.5278}}].

\bibitem{Kang:2015msa}
Z.-B. Kang, A.~Prokudin, P.~Sun and F.~Yuan, \emph{{Extraction of Quark
  Transversity Distribution and Collins Fragmentation Functions with QCD
  Evolution}}, \href{https://doi.org/10.1103/PhysRevD.93.014009}{\emph{Phys.
  Rev.} {\bfseries D93} (2016) 014009}
  [\href{https://arxiv.org/abs/1505.05589}{{\ttfamily 1505.05589}}].

\bibitem{Anselmino:2015sxa}
M.~Anselmino, M.~Boglione, U.~D'Alesio, J.~O. Gonzalez~Hernandez, S.~Melis,
  F.~Murgia et~al., \emph{{Collins functions for pions from SIDIS and new
  $e^+e^-$ data: a first glance at their transverse momentum dependence}},
  \href{https://doi.org/10.1103/PhysRevD.92.114023}{\emph{Phys. Rev.}
  {\bfseries D92} (2015) 114023}
  [\href{https://arxiv.org/abs/1510.05389}{{\ttfamily 1510.05389}}].

\bibitem{Bacchetta:2011ip}
A.~Bacchetta, A.~Courtoy and M.~Radici, \emph{{First glances at the
  transversity parton distribution through dihadron fragmentation functions}},
  \href{https://doi.org/10.1103/PhysRevLett.107.012001}{\emph{Phys.Rev.Lett.}
  {\bfseries 107} (2011) 012001}
  [\href{https://arxiv.org/abs/1104.3855}{{\ttfamily 1104.3855}}].

\bibitem{Pisano:2015wnq}
S.~Pisano and M.~Radici, \emph{{Di-hadron fragmentation and mapping of the
  nucleon structure}},
  \href{https://doi.org/10.1140/epja/i2016-16155-5}{\emph{Eur. Phys. J.}
  {\bfseries A52} (2016) 155}
  [\href{https://arxiv.org/abs/1511.03220}{{\ttfamily 1511.03220}}].

\bibitem{Airapetian:2004tw}
{\scshape HERMES Collaboration} collaboration, A.~Airapetian et~al.,
  \emph{{Single-spin asymmetries in semi-inclusive deep-inelastic scattering on
  a transversely polarized hydrogen target}},
  \href{https://doi.org/10.1103/PhysRevLett.94.012002}{\emph{Phys.Rev.Lett.}
  {\bfseries 94} (2005) 012002}
  [\href{https://arxiv.org/abs/hep-ex/0408013}{{\ttfamily hep-ex/0408013}}].

\bibitem{Airapetian:2008sk}
{\scshape HERMES Collaboration} collaboration, A.~Airapetian et~al.,
  \emph{{Evidence for a Transverse Single-Spin Asymmetry in Leptoproduction of
  pi+pi- Pairs}},
  \href{https://doi.org/10.1088/1126-6708/2008/06/017}{\emph{JHEP} {\bfseries
  06} (2008) 017} [\href{https://arxiv.org/abs/0803.2367}{{\ttfamily
  0803.2367}}].

\bibitem{Adolph:2012nw}
{\scshape COMPASS Collaboration} collaboration, C.~Adolph et~al.,
  \emph{{Transverse spin effects in hadron-pair production from semi-inclusive
  deep inelastic scattering}},
  \href{https://doi.org/10.1016/j.physletb.2012.05.015}{\emph{Phys.Lett.}
  {\bfseries B713} (2012) 10}
  [\href{https://arxiv.org/abs/1202.6150}{{\ttfamily 1202.6150}}].

\bibitem{Adolph:2014fjw}
{\scshape COMPASS Collaboration} collaboration, C.~Adolph et~al., \emph{{A
  high-statistics measurement of transverse spin effects in dihadron production
  from muon-proton semi-inclusive deep-inelastic scattering}},
  \href{https://doi.org/10.1016/j.physletb.2014.06.080}{\emph{Phys.Lett.}
  {\bfseries B736} (2014) 124}
  [\href{https://arxiv.org/abs/1401.7873}{{\ttfamily 1401.7873}}].

\bibitem{Meissner:2008yf}
S.~Meissner and A.~Metz, \emph{{Partonic pole matrix elements for
  fragmentation}},
  \href{https://doi.org/10.1103/PhysRevLett.102.172003}{\emph{Phys. Rev. Lett.}
  {\bfseries 102} (2009) 172003}
  [\href{https://arxiv.org/abs/0812.3783}{{\ttfamily 0812.3783}}].

\bibitem{Gamberg:2010uw}
L.~P. Gamberg, A.~Mukherjee and P.~J. Mulders, \emph{{A model independent
  analysis of gluonic pole matrix elements and universality of TMD
  fragmentation functions}},
  \href{https://doi.org/10.1103/PhysRevD.83.071503}{\emph{Phys. Rev.}
  {\bfseries D83} (2011) 071503}
  [\href{https://arxiv.org/abs/1010.4556}{{\ttfamily 1010.4556}}].

\bibitem{Bacchetta:2003vn}
A.~Bacchetta and M.~Radici, \emph{{Two hadron semiinclusive production
  including subleading twist}},
  \href{https://doi.org/10.1103/PhysRevD.69.074026}{\emph{Phys.Rev.} {\bfseries
  D69} (2004) 074026} [\href{https://arxiv.org/abs/hep-ph/0311173}{{\ttfamily
  hep-ph/0311173}}].

\bibitem{Collins:2002kn}
J.~C. Collins, \emph{{Leading twist single transverse-spin asymmetries:
  Drell-Yan and deep inelastic scattering}},
  \href{https://doi.org/10.1016/S0370-2693(02)01819-1}{\emph{Phys.Lett.}
  {\bfseries B536} (2002) 43}
  [\href{https://arxiv.org/abs/hep-ph/0204004}{{\ttfamily hep-ph/0204004}}].

\bibitem{Boer:2010ya}
D.~Boer, Z.-B. Kang, W.~Vogelsang and F.~Yuan, \emph{{Test of the Universality
  of Naive-time-reversal-odd Fragmentation Functions}},
  \href{https://doi.org/10.1103/PhysRevLett.105.202001}{\emph{Phys. Rev. Lett.}
  {\bfseries 105} (2010) 202001}
  [\href{https://arxiv.org/abs/1008.3543}{{\ttfamily 1008.3543}}].

\bibitem{Matevosyan:2017liq}
H.~H. Matevosyan, A.~Kotzinian and A.~W. Thomas, \emph{{Accessing quark
  helicity through dihadron studies}},
  \href{https://doi.org/10.1103/PhysRevLett.120.252001}{\emph{Phys. Rev. Lett.}
  {\bfseries 120} (2018) 252001}
  [\href{https://arxiv.org/abs/1712.06384}{{\ttfamily 1712.06384}}].

\bibitem{Bacchetta:2002ux}
A.~Bacchetta and M.~Radici, \emph{{Partial wave analysis of two hadron
  fragmentation functions}},
  \href{https://doi.org/10.1103/PhysRevD.67.094002}{\emph{Phys. Rev.}
  {\bfseries D67} (2003) 094002}
  [\href{https://arxiv.org/abs/hep-ph/0212300}{{\ttfamily hep-ph/0212300}}].

\bibitem{Boer:2003ya}
D.~Boer, R.~Jakob and M.~Radici, \emph{{Interference fragmentation functions in
  electron positron annihilation}},
  \href{https://doi.org/10.1103/PhysRevD.67.094003}{\emph{Phys.Rev.} {\bfseries
  D67} (2003) 094003} [\href{https://arxiv.org/abs/hep-ph/0302232}{{\ttfamily
  hep-ph/0302232}}].

\bibitem{Matevosyan:2018icf}
H.~H. Matevosyan, A.~Bacchetta, D.~Boer, A.~Courtoy, A.~Kotzinian, M.~Radici
  et~al., \emph{{Semi-inclusive production of two back-to-back hadron pairs in
  $e^+e^-$ annihilation revisited}},
  \href{https://doi.org/10.1103/PhysRevD.97.074019}{\emph{Phys. Rev.}
  {\bfseries D97} (2018) 074019}
  [\href{https://arxiv.org/abs/1802.01578}{{\ttfamily 1802.01578}}].

\bibitem{Boer:1997mf}
D.~Boer, R.~Jakob and P.~Mulders, \emph{{Asymmetries in polarized hadron
  production in e+ e- annihilation up to order 1/Q}},
  \href{https://doi.org/10.1016/S0550-3213(97)00456-2}{\emph{Nucl.Phys.}
  {\bfseries B504} (1997) 345}
  [\href{https://arxiv.org/abs/hep-ph/9702281}{{\ttfamily hep-ph/9702281}}].

\bibitem{Boer:1997qn}
D.~Boer, R.~Jakob and P.~J. Mulders, \emph{{Leading asymmetries in two hadron
  production in e+ e- annihilation at the Z pole}},
  \href{https://doi.org/10.1016/S0370-2693(98)00136-1}{\emph{Phys. Lett.}
  {\bfseries B424} (1998) 143}
  [\href{https://arxiv.org/abs/hep-ph/9711488}{{\ttfamily hep-ph/9711488}}].

\bibitem{Bianconi:1999cd}
A.~Bianconi, S.~Boffi, R.~Jakob and M.~Radici, \emph{{Two hadron interference
  fragmentation functions. Part 1. General framework}},
  \href{https://doi.org/10.1103/PhysRevD.62.034008}{\emph{Phys.Rev.} {\bfseries
  D62} (2000) 034008} [\href{https://arxiv.org/abs/hep-ph/9907475}{{\ttfamily
  hep-ph/9907475}}].

\bibitem{Radici:2001na}
M.~Radici, R.~Jakob and A.~Bianconi, \emph{{Accessing transversity with
  interference fragmentation functions}},
  \href{https://doi.org/10.1103/PhysRevD.65.074031}{\emph{Phys.Rev.} {\bfseries
  D65} (2002) 074031} [\href{https://arxiv.org/abs/hep-ph/0110252}{{\ttfamily
  hep-ph/0110252}}].

\bibitem{Boer:2008fr}
D.~Boer, \emph{{Angular dependences in inclusive two-hadron production at
  BELLE}}, \href{https://doi.org/10.1016/j.nuclphysb.2008.06.011}{\emph{Nucl.
  Phys.} {\bfseries B806} (2009) 23}
  [\href{https://arxiv.org/abs/0804.2408}{{\ttfamily 0804.2408}}].

\bibitem{Mulders:1995dh}
P.~Mulders and R.~Tangerman, \emph{{The Complete tree level result up to order
  1/Q for polarized deep inelastic leptoproduction}},
  \href{https://doi.org/10.1016/0550-3213(95)00632-X,
  10.1016/0550-3213(95)00632-X}{\emph{Nucl.Phys.} {\bfseries B461} (1996) 197}
  [\href{https://arxiv.org/abs/hep-ph/9510301}{{\ttfamily hep-ph/9510301}}].

\bibitem{Seidl:2015lla}
{\scshape Belle} collaboration, R.~Seidl et~al., \emph{{Inclusive cross
  sections for pairs of identified light charged hadrons and for single protons
  in $e^+e^-$ at $\sqrt{s}=$ 10.58 GeV}},
  \href{https://doi.org/10.1103/PhysRevD.92.092007}{\emph{Phys. Rev.}
  {\bfseries D92} (2015) 092007}
  [\href{https://arxiv.org/abs/1509.00563}{{\ttfamily 1509.00563}}].

\bibitem{deFlorian:2014xna}
D.~de~Florian, R.~Sassot, M.~Epele, R.~J. Hern‡ndez-Pinto and M.~Stratmann,
  \emph{{Parton-to-Pion Fragmentation Reloaded}},
  \href{https://doi.org/10.1103/PhysRevD.91.014035}{\emph{Phys. Rev.}
  {\bfseries D91} (2015) 014035}
  [\href{https://arxiv.org/abs/1410.6027}{{\ttfamily 1410.6027}}].

\bibitem{Hirai:2016loo}
M.~Hirai, H.~Kawamura, S.~Kumano and K.~Saito, \emph{{Impacts of B-factory
  measurements on determination of fragmentation functions from
  electron-positron annihilation data}},
  \href{https://doi.org/10.1093/ptep/ptw154}{\emph{PTEP} {\bfseries 2016}
  (2016) 113B04} [\href{https://arxiv.org/abs/1608.04067}{{\ttfamily
  1608.04067}}].

\bibitem{Sato:2016wqj}
N.~Sato, J.~J. Ethier, W.~Melnitchouk, M.~Hirai, S.~Kumano and A.~Accardi,
  \emph{{First Monte Carlo analysis of fragmentation functions from
  single-inclusive $e^+ e^-$ annihilation}},
  \href{https://doi.org/10.1103/PhysRevD.94.114004}{\emph{Phys. Rev.}
  {\bfseries D94} (2016) 114004}
  [\href{https://arxiv.org/abs/1609.00899}{{\ttfamily 1609.00899}}].

\bibitem{Ethier:2017zbq}
J.~J. Ethier, N.~Sato and W.~Melnitchouk, \emph{{First simultaneous extraction
  of spin-dependent parton distributions and fragmentation functions from a
  global QCD analysis}},
  \href{https://doi.org/10.1103/PhysRevLett.119.132001}{\emph{Phys. Rev. Lett.}
  {\bfseries 119} (2017) 132001}
  [\href{https://arxiv.org/abs/1705.05889}{{\ttfamily 1705.05889}}].

\bibitem{Bertone:2017tyb}
{\scshape NNPDF} collaboration, V.~Bertone, S.~Carrazza, N.~P. Hartland, E.~R.
  Nocera and J.~Rojo, \emph{{A determination of the fragmentation functions of
  pions, kaons, and protons with faithful uncertainties}},
  \href{https://doi.org/10.1140/epjc/s10052-017-5088-y}{\emph{Eur. Phys. J.}
  {\bfseries C77} (2017) 516}
  [\href{https://arxiv.org/abs/1706.07049}{{\ttfamily 1706.07049}}].

\bibitem{Ito:2009zc}
T.~Ito, W.~Bentz, I.~C. Cloet, A.~W. Thomas and K.~Yazaki, \emph{{The NJL-jet
  model for quark fragmentation functions}},
  \href{https://doi.org/10.1103/PhysRevD.80.074008}{\emph{Phys. Rev.}
  {\bfseries D80} (2009) 074008}
  [\href{https://arxiv.org/abs/0906.5362}{{\ttfamily 0906.5362}}].

\bibitem{Matevosyan:2011zza}
H.~H. Matevosyan, A.~W. Thomas and W.~Bentz, \emph{{Analyzing unfavored
  fragmentation functions using NJL-jet model}},
  \href{https://doi.org/10.1063/1.3647165}{\emph{AIP Conf.Proc.} {\bfseries
  1374} (2011) 387}.

\bibitem{Matevosyan:2011ey}
H.~H. Matevosyan, A.~W. Thomas and W.~Bentz, \emph{{Monte Carlo Simulations of
  Hadronic Fragmentation Functions using NJL-Jet Model}},
  \href{https://doi.org/10.1103/PhysRevD.83.114010,
  10.1103/PhysRevD.86.059904}{\emph{Phys.Rev.} {\bfseries D83} (2011) 114010}
  [\href{https://arxiv.org/abs/1103.3085}{{\ttfamily 1103.3085}}].

\bibitem{Matevosyan:2013aka}
H.~H. Matevosyan, A.~W. Thomas and W.~Bentz, \emph{{Dihadron Fragmentation
  Functions within the NJL-jet Model}},
  \href{https://doi.org/10.1103/PhysRevD.88.094022}{\emph{Phys.Rev.} {\bfseries
  D88} (2013) 094022} [\href{https://arxiv.org/abs/1310.1917}{{\ttfamily
  1310.1917}}].

\bibitem{Boer:1997nt}
D.~Boer and P.~J. Mulders, \emph{{Time reversal odd distribution functions in
  leptoproduction}},
  \href{https://doi.org/10.1103/PhysRevD.57.5780}{\emph{Phys. Rev.} {\bfseries
  D57} (1998) 5780} [\href{https://arxiv.org/abs/hep-ph/9711485}{{\ttfamily
  hep-ph/9711485}}].

\bibitem{Berestetsky:1982aq}
V.~B. Berestetskii, E.~M. Lifshitz and L.~P. Pitaevskii, \emph{{QUANTUM
  ELECTRODYNAMICS}}, vol.~4 of \emph{Course of Theoretical Physics}. Pergamon
  Press, Oxford, 1982.

\bibitem{Matevosyan:2016fwi}
H.~H. Matevosyan, A.~Kotzinian and A.~W. Thomas, \emph{{Monte Carlo
  Implementation of Polarized Hadronization}},
  \href{https://doi.org/10.1103/PhysRevD.95.014021}{\emph{Phys. Rev.}
  {\bfseries D95} (2017) 014021}
  [\href{https://arxiv.org/abs/1610.05624}{{\ttfamily 1610.05624}}].

\bibitem{Cronin:1963zb}
J.~W. Cronin and O.~E. Overseth, \emph{{Measurement of the decay parameters of
  the Lambda0 particle}},
  \href{https://doi.org/10.1103/PhysRev.129.1795}{\emph{Phys. Rev.} {\bfseries
  129} (1963) 1795}.

\bibitem{Beringer:1900zz}
{\scshape Particle Data Group} collaboration, J.~Beringer et~al., \emph{{Review
  of Particle Physics (RPP)}},
  \href{https://doi.org/10.1103/PhysRevD.86.010001}{\emph{Phys.Rev.} {\bfseries
  D86} (2012) 010001}.

\bibitem{Lee:1957qs}
T.~D. Lee and C.-N. Yang, \emph{{General Partial Wave Analysis of the Decay of
  a Hyperon of Spin 1/2}},
  \href{https://doi.org/10.1103/PhysRev.108.1645}{\emph{Phys. Rev.} {\bfseries
  108} (1957) 1645}.

\bibitem{Bigi:1976qt}
I.~I.~Y. Bigi, \emph{{Transfer of Polarization from the Initial to the Final
  State in Deep Inelastic Lepton Scattering}},
  \href{https://doi.org/10.1007/BF02730452}{\emph{Nuovo Cim.} {\bfseries A41}
  (1977) 43}.

\bibitem{Bigi:1977qe}
I.~I.~Y. Bigi, \emph{{Some Quantitative Estimates About Final State
  Polarization in Deep Inelastic Lepton-Nucleon Scattering}},
  \href{https://doi.org/10.1007/BF02730200}{\emph{Nuovo Cim.} {\bfseries A41}
  (1977) 581}.

\bibitem{Augustin:1978wf}
J.~E. Augustin and F.~M. Renard, \emph{{How to Measure Quark Helicities in $e^+
  e^- \to$ Hadrons}},
  \href{https://doi.org/10.1016/0550-3213(80)90269-2}{\emph{Nucl. Phys.}
  {\bfseries B162} (1980) 341}.

\bibitem{Baldracchini:1980uq}
F.~Baldracchini, N.~S. Craigie, V.~Roberto and M.~Socolovsky, \emph{{A Survey
  of Polarization Asymmetries Predicted by {QCD}}},
  \href{https://doi.org/10.1002/prop.19810291102}{\emph{Fortsch. Phys.}
  {\bfseries 30} (1981) 505}.

\bibitem{Buskulic:1996vb}
{\scshape ALEPH} collaboration, D.~Buskulic et~al., \emph{{Measurement of
  Lambda polarization from Z decays}},
  \href{https://doi.org/10.1016/0370-2693(96)00300-0}{\emph{Phys. Lett.}
  {\bfseries B374} (1996) 319}.

\bibitem{Astier:2000ax}
{\scshape NOMAD} collaboration, P.~Astier et~al., \emph{{Measurement of the
  Lambda polarization in nu/mu charged current interactions in the NOMAD
  experiment}},
  \href{https://doi.org/10.1016/S0550-3213(00)00503-4}{\emph{Nucl. Phys.}
  {\bfseries B588} (2000) 3}.

\bibitem{Alekseev:2009ab}
{\scshape COMPASS} collaboration, M.~Alekseev et~al., \emph{{Measurement of the
  Longitudinal Spin Transfer to Lambda and Anti-Lambda Hyperons in Polarised
  Muon DIS}}, \href{https://doi.org/10.1140/epjc/s10052-009-1143-7}{\emph{Eur.
  Phys. J.} {\bfseries C64} (2009) 171}
  [\href{https://arxiv.org/abs/0907.0388}{{\ttfamily 0907.0388}}].

\bibitem{Ellis:1995fc}
J.~R. Ellis, D.~Kharzeev and A.~Kotzinian, \emph{{The Proton spin puzzle and
  lambda polarization in deep inelastic scattering}},
  \href{https://doi.org/10.1007/s002880050048}{\emph{Z. Phys.} {\bfseries C69}
  (1996) 467} [\href{https://arxiv.org/abs/hep-ph/9506280}{{\ttfamily
  hep-ph/9506280}}].

\bibitem{Kotzinian:1997vd}
A.~Kotzinian, A.~Bravar and D.~von Harrach, \emph{{Lambda and anti-Lambda
  polarization in lepton induced processes}},
  \href{https://doi.org/10.1007/s100520050142}{\emph{Eur. Phys. J.} {\bfseries
  C2} (1998) 329} [\href{https://arxiv.org/abs/hep-ph/9701384}{{\ttfamily
  hep-ph/9701384}}].

\bibitem{Ellis:2002zv}
J.~R. Ellis, A.~Kotzinian and D.~V. Naumov, \emph{{Intrinsic polarized
  strangeness and Lambda0 polarization in deep inelastic production}},
  \href{https://doi.org/10.1140/epjc/s2002-01025-2}{\emph{Eur. Phys. J.}
  {\bfseries C25} (2002) 603}
  [\href{https://arxiv.org/abs/hep-ph/0204206}{{\ttfamily hep-ph/0204206}}].

\bibitem{Ellis:2007ig}
J.~R. Ellis, A.~Kotzinian, D.~Naumov and M.~Sapozhnikov, \emph{{Longitudinal
  Polarization of Lambda and anti-Lambda Hyperons in Lepton-Nucleon
  Deep-Inelastic Scattering}},
  \href{https://doi.org/10.1140/epjc/s10052-007-0381-9}{\emph{Eur. Phys. J.}
  {\bfseries C52} (2007) 283}
  [\href{https://arxiv.org/abs/hep-ph/0702222}{{\ttfamily hep-ph/0702222}}].

\bibitem{Abdesselam:2016nym}
{\scshape Belle} collaboration, A.~Abdesselam et~al., \emph{{Observation of
  Transverse $\Lambda/\bar{\Lambda}$ Hyperon Polarization in $e^+e^-$
  Annihilation at Belle}},  \href{https://arxiv.org/abs/1611.06648}{{\ttfamily
  1611.06648}}.

\bibitem{Guan:2018ckx}
{\scshape Belle} collaboration, Y.~Guan et~al., \emph{{Observation of
  Transverse $\Lambda/\bar{\Lambda}$ Hyperon Polarization in $e^+e^-$
  Annihilation at Belle}},  \href{https://arxiv.org/abs/1808.05000}{{\ttfamily
  1808.05000}}.

\bibitem{Pitonyak:2013dsu}
D.~Pitonyak, M.~Schlegel and A.~Metz, \emph{{Polarized hadron pair production
  from electron-positron annihilation}},
  \href{https://doi.org/10.1103/PhysRevD.89.054032}{\emph{Phys. Rev.}
  {\bfseries D89} (2014) 054032}
  [\href{https://arxiv.org/abs/1310.6240}{{\ttfamily 1310.6240}}].

\bibitem{Matevosyan:2018dea}
H.~H. Matevosyan, \emph{{Accessing quark helicity in $e^+e^-$ and SIDIS via
  dihadron correlations}},  \href{https://arxiv.org/abs/1807.11485}{{\ttfamily
  1807.11485}}.

\end{thebibliography}\endgroup

\end{document}